\documentclass{article}

\usepackage{xcolor}

\usepackage{soul}

\usepackage{subfig}

\usepackage{fullpage}
\usepackage{amssymb,latexsym,amsmath}     
\usepackage{graphicx}
\usepackage{dsfont}
\usepackage{amsthm}
\usepackage{color}
\usepackage{float}
\usepackage{grffile}
\usepackage{blindtext}
\usepackage[T1]{fontenc}
\usepackage[utf8]{inputenc}
\usepackage{graphicx}
\usepackage{dsfont}
\usepackage{comment}
\usepackage{amssymb}
\usepackage{dsfont}
\usepackage{latexsym}
\usepackage{dcolumn}
\usepackage{caption}
\usepackage{amsmath}
\usepackage{epsf}
\usepackage{enumerate}
\usepackage{url} 
\usepackage{caption}
\usepackage{tikz}
\usepackage{sectsty}
\usepackage{siunitx}
\usepackage{svg}

\usepackage{graphicx}
\usepackage{dcolumn}
\usepackage{bm}
\usepackage{mathtools}

\usepackage{booktabs}
\usepackage{blkarray}

\usepackage{authblk}
\usepackage{lineno}

\title{Calculating the precision of tilt-to-length coupling estimation and noise subtraction in LISA using Fisher information} 

\begin{document}

\author[1]{Daniel George}
\author[2]{Jose Sanjuan}
\author[1]{Paul Fulda}
\author[1]{Guido Mueller}
\affil[1]{Department of Physics, University of Florida, PO Box 118440, Gainesville, FL 32611-8440, USA}
\affil[2]{Department of Aerospace Engineering, Texas A\&M University, 701 H.R. Bright Bldg. 3141, College Station, TX 77843, USA}

\date{}

\maketitle

\begin{abstract}
    Tilt-to-length (TTL) noise from angular jitter in LISA is projected to be the dominant noise source in the milli-Hertz band unless corrected in post-processing. The correction is only possible after removing the overwhelming laser phase noise using time-delay interferometry (TDI). 
    We present here a frequency domain model that describes the effect of angular motion of all three spacecraft on the interferometric signals after propagating through TDI. We then apply a Fisher information matrix analysis to this model to calculate the minimum uncertainty with which TTL coupling coefficients may be estimated.
    Furthermore, we show the impact of these uncertainties on the residual TTL noise in the gravitational wave readout channel, and compare it to the impact of the angular witness sensors' readout noise. 
     We show that the residual TTL noise post-subtraction in the TDI variables for a case using the LISA angular jitter requirement and integration time of one day is limited to the 8\,pm/$\sqrt{\rm Hz}$ level by angular sensing noise.
     However, using a more realistic model for the angular jitter we find that the TTL coupling uncertainties are 70 times larger,
     and the noise subtraction is limited by these uncertainties to the 14\,pm/$\sqrt{\rm Hz}$ level. 

\end{abstract}

\section{Introduction \label{sec:intro}}
    
\begin{figure}[h!]
\begin{center}
    \includegraphics[width=0.5\linewidth]{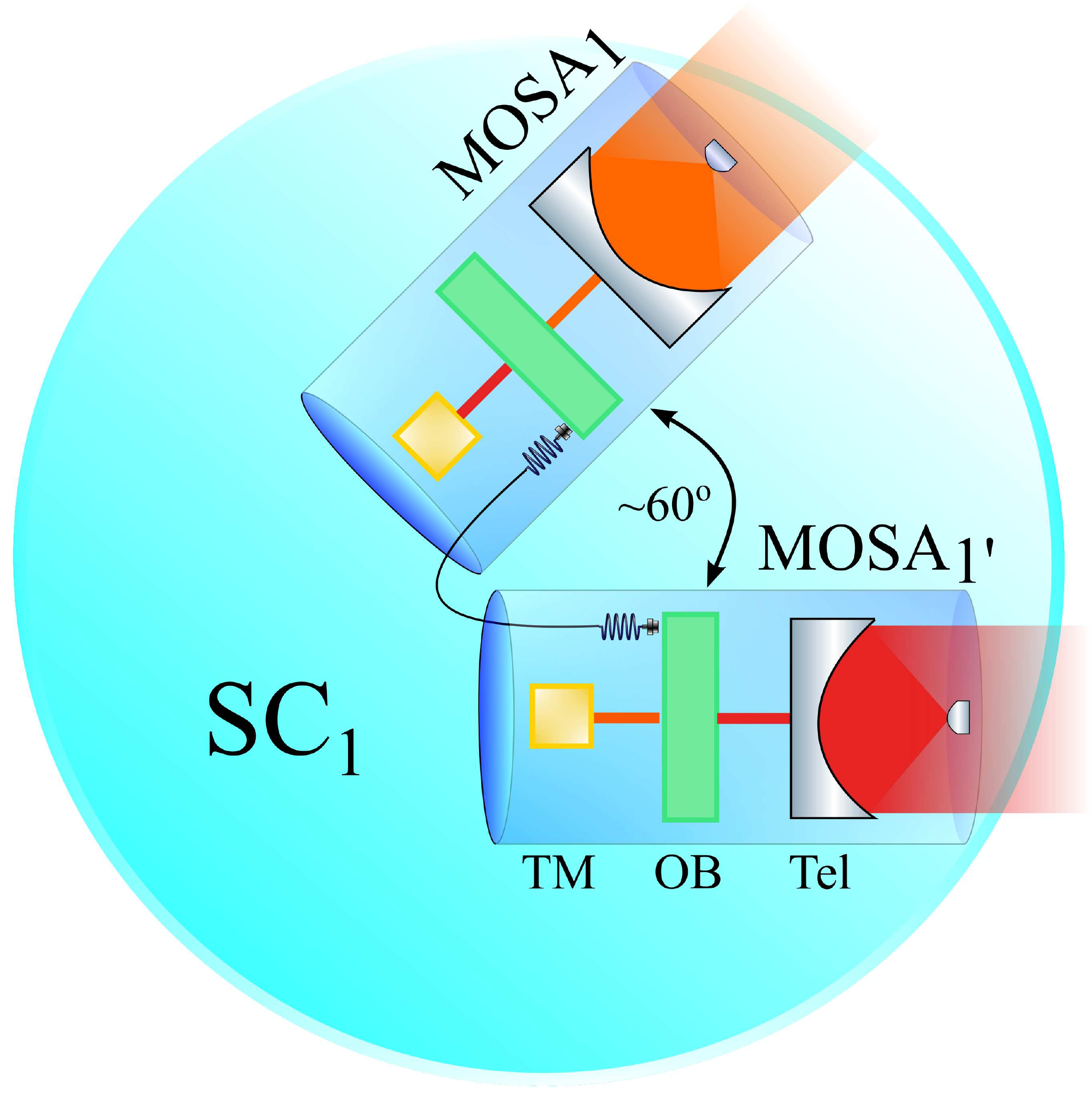}
\end{center}
\caption{Simplified schematic of the movable optical sub-assemblies (MOSAs) on a single SC. TM: test mass; OB: optical bench, connected through the backlink fiber; Tel: telescope. The telescope transmits the local beam with magnification of around 300 (134 telescope magnification and about 2.2 from the OB), and simultaneously receives light from the far SCs. \label{fig.LISA_MOSA}}
\end{figure}

The Laser Interferometer Space Antenna (LISA)~\cite{LISA_1, LISA_2}, the space-based gravitational wave (GW) detector led by the European Space Agency (ESA) with contributions from the National Aeronautics and Space Administration (NASA), is scheduled for launch in the 2030s and will complement ground-based detectors with sensitivity in the low-frequency range, 0.1\,mHz to 1\,Hz~\cite{shutz_1997, LISA_sources, LISA_sources_2}. LISA consists of three spacecraft (SC) in an almost equilateral triangular formation with sides of 2.5\,Gm. Each SC has two movable optical sub-assemblies (MOSA) that include a gravitational reference system (GRS), an optical bench (OB) and a telescope; Figure~\ref{fig.LISA_MOSA} shows a very simplified drawing of two of the six MOSAs while the constellation is sketched in Fig.~\ref{fig.LISA_full}. The telescope will simultaneously receive light from and also transmit light to the far spacecraft. It exchanges light bidirectionally with the OB. Each GRS houses one free falling test mass (TM), which acts as one end-point for the critical TM to TM distance. However, the entire distance is measured on the OB in three steps: local TM to local OB, local OB to far OB, and then far OB to far TM. 

The science interferometer measures distances between the OBs connected along an optical arm in the constellation. This measurement will be contaminated by additional length noise from OB  motion which is attached to its MOSA and the local SC. The distance from the TM to the OB is measured by the local interferometer. To subtract out the OB displacement noise, we use~\cite{TDI_otto}’s method of subtracting the TM-OB measurement from the science interferometer. We then apply time-delay interferometry (TDI) 2.0~\cite{TDI_1, TDI_2, TDI_3, TDI_4, TDI_living_reviews} to suppress laser phase noise (LPN) by considering the simple Michelson variables $X$, $Y$, and $Z$, and assume the arm-lengths are perfectly known. 

During the operational run of LISA Pathfinder (LPF), TTL dominated the sensitivity curve between 20 and 200\,mHz with noise levels at the pm/$\sqrt{\rm Hz}$ level before they were subtracted in post-processing~\cite{LPF_TTL, LPF_DFACS, armano_et_al_2019}. A similar approach will be taken in LISA to correct for TTL in the local interferometer.  In this paper, we focus on TTL in the science interferometer where the effects are magnified by the telescope magnification~\cite{TTL_telescope_mag} of around 300 (134 telescope magnification and about 2.2 from the OB) and which are a critical concern for the LISA mission.

The noise levels induced by TTL will exceed the noise floor unless mitigating actions are taken. These include precise SC attitude control, active alignment corrections in orbit, and subtraction in post-processing. Post-processing requires to measure the angular motion using LISA's differential wavefront sensing (DWS) system~\cite{Morrison:94} in-situ, then scale this data with the proper TTL calibration factor before the spurious length changes can be subtracted. The TTL calibration factors need to be measured regularly to account for slow changes over time, with high enough accuracy before TTL subtraction can be successfully applied. Furthermore, knowledge of the individual TTL coefficients can ease alignment processes in-orbit and identify issues in specific subsystems, e.g., whether a telescope exhibits significantly higher WFE, or perhaps a large misalignment in a MOSA. 

The TTL coupling coefficients will be measured using correlations between the DWS signals and the length signals after TDI has been applied. The accuracy will depend on the angular jitter itself ---in general larger jitter helps to determine coefficients better, but also requires more accurate coefficients to subtract to a given residual level. We will consider two cases: case A will use the LISA requirement jitter and length noise, while case B utilizes a jitter and length noise estimates from LISASim~\cite{lisaSIM}.

TTL coefficients depend on multiple alignment degrees of freedom and require redoing calibration regularly. In this paper, we answer the question of how well each component of TTL noise can be estimated, and consequently, subtracted within the framework of TDI. This is achieved by building a quantitative model of how TTL effects couple into the interferometric signal. One can then calculate the expected errors for each coefficient by calculating its Fisher information~\cite{FIM, FIM_2,PhysRevD.77.042001}. Based on the results we obtain, we can also estimate the required integration times to calibrate TTL sufficiently.

We calculate the lower limits on the TTL coefficients using the Cram\'{e}r-Rao bound~\cite{C_Rao, C_Rao_2} and then compare them with time-domain Monte Carlo simulations where we find good agreement. This paper is organized as follows: section~\ref{sec:the.model} describes the model we use to quantify TTL noise. Section~\ref{sec:FIM.in.TDI} goes through the Fisher information analysis of TTL noise in TDI. We end by displaying the results in section~\ref{sec:results}, followed by the discussion in section~\ref{sec:final}.

\section{The Model \label{sec:the.model}}

Angular jitters in an interferometric measurement may result in spurious longitudinal displacements termed TTL noise, due to geometrical and non-geometrical contributions~\cite{PhysRevApplied.14.014030,10.1088/2040-8986/ac675e,Weaver:20,hasselmann_brugger_vogel_fitzsimsons_johann_heinzel_weise_sell_2021}. For a sufficiently small dynamic range, the effects for each degree of freedom can be described by a linear relationship between angle and displacement characterized by the TTL coupling coefficients in units of m/rad. 
    
\begin{figure}[h!]
\begin{center}
    \includegraphics[width=0.75\linewidth]{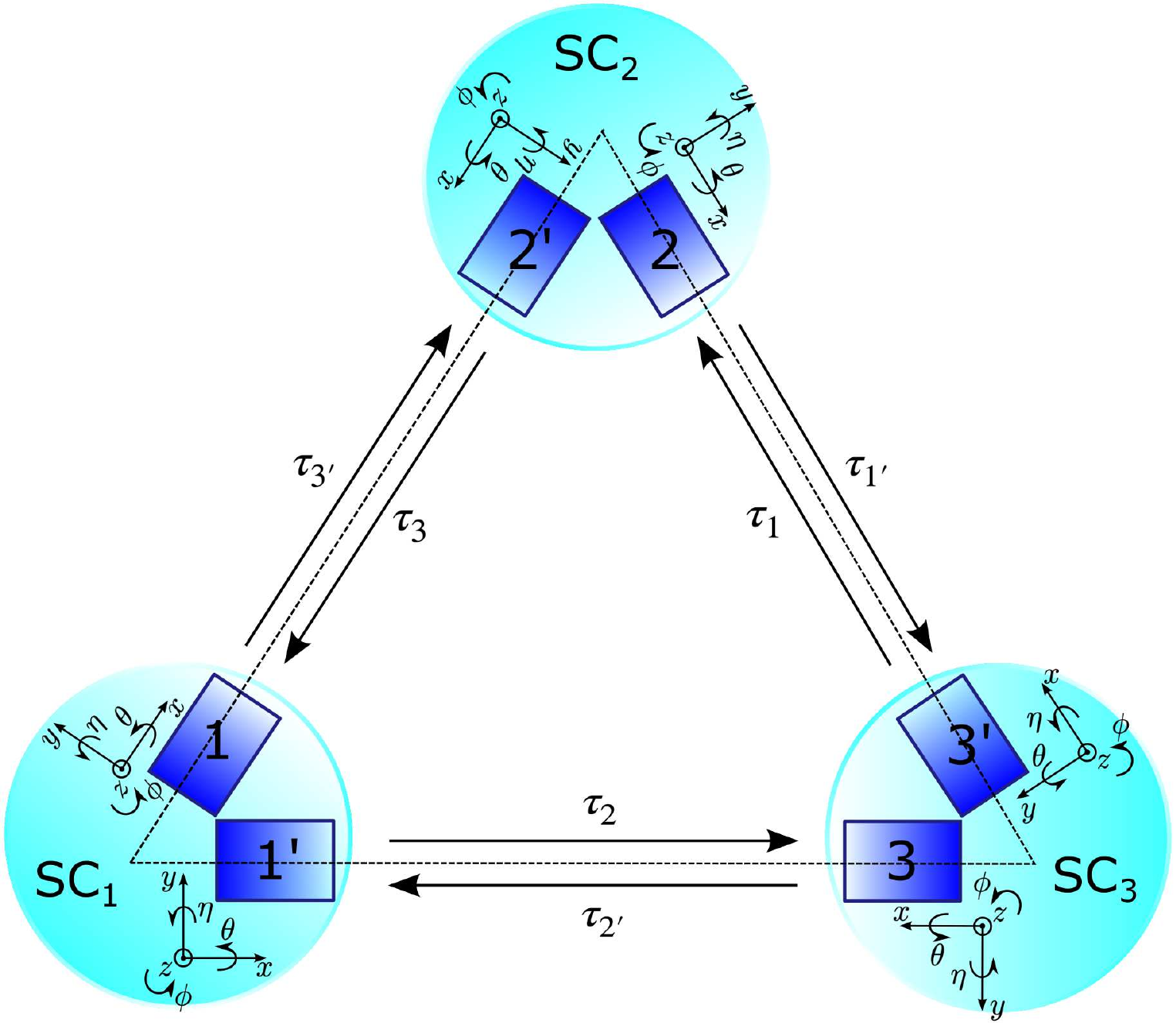}
\end{center}
\caption{LISA constellation and reference frames. The blue rectangles represent the MOSAs with the primed and non-primed nomenclature as in the literature~\cite{TDI_living_reviews}, with their respective coordinate axes. Yaw ($\phi$) is positive for counterclockwise rotation around the z-axis; pitch ($\eta$) is positive for counterclockwise rotation around the y-axis; roll ($\theta$) is positive for counterclockwise rotation around the x-axis. Every SC has two MOSAs with an opening angle of roughly $60^{\mathrm o}$, forming a roughly equilateral triangle constellation. Each primed MOSA forms a science interferometer with a non-primed MOSA. The $i^{\rm th}$ delay, $\tau_{i}$, along an arm corresponds to the opposite SC$_{i}$, with counterclockwise (clockwise) light travel paths being denoted by primed (non-primed) delays. \label{fig.LISA_full}}
\end{figure}

We assign a unique TTL coefficient to each MOSA for the angular degrees of freedom of interest, i.e., yaw ($\phi$) and pitch ($\eta$); Figure~\ref{fig.LISA_full} displays the MOSAs, the reference axis, and the nomenclature we use in labeling the MOSAs and delays. If a MOSA is jittering, it will impart TTL noise to the received (RX) beam measured locally, and to the transmitted (TX) beam, measured on the far SC. As seen in Fig.~\ref{fig.LISA_full}, we will have 24 TTL coefficients from 6 MOSAs $\times$ 2 angular degrees of freedom $\times$ 2 directions (RX, TX).

To cancel the OB motion noise present in the science interferometer signals, we utilize the method in \cite{TDI_otto} of constructing a new variable $\xi_i$ that subtracts the local $\varepsilon_i$ and reference $\tau_i$ interferometers from the science $s_i$ interferometer (see Eqs.~\ref{eq.xi_signals1}-\ref{eq.xi_signals2}), as we are mainly considered with TTL noise from the science interferometer: 
\begin{eqnarray}
    \xi_{i} &=& {\mathbf k}^{\rm RX}_{i}{\mathbf \Psi}_{i} + {\mathbf k}^{\rm RX}_{j'}{\mathbf \Psi}_{j':k} + {\rm noise_i}  \label{eq.xi} \\
    \xi_{i'} &=& {\mathbf k}^{\rm RX}_{i'}{\mathbf \Psi}_{i'} + {\mathbf k}^{\rm RX}_{k}{\mathbf \Psi}_{k:j'} + {\rm noise_{i'}}  \label{eq.xip} 
\end{eqnarray}
where the noise comes from the interferometer, residual test mass acceleration, the GW background, and LPN which will be removed in TDI 2.0. $(c, d) = {\mathbf k}$ are the TTL coefficients associated with the yaw and pitch degrees of freedom respectively. These angular degrees of freedom are measured by the DWS signals $\mathbf{\Psi} = (\phi, \eta)$, where $\phi$ and $\eta$ are the DWS signals comprising of the angular motion and the associated readout noise. The delay nomenclature follows the literature where, e.g., ${\mathbf \Psi}_{j':k} \equiv {\mathbf \Psi}_{j'}(t - \tau_k)$. The coefficients are defined to be positive when for a positive rotation the optical path length increases. The angles are defined with respect to the incoming flat wave-front and described in their own reference frame: $\phi$ positive means counterclockwise rotation, $\eta$ positive means also counterclockwise when looking towards $x$ --- see Fig.~\ref{fig.LISA_full}. All the six phasemeter signals can be obtained with cyclic permutations of the indices $ijk$=123, 231, and 312.
    
To keep the heterodyne signals within the bandwidth of the photodetectors (5-20\,MHz), the so-called primary-secondary laser transponder configuration must be adopted: the laser of one MOSA is set as the primary one, while the remaining lasers are frequency offset phase-locked to it. We adopt the same laser transponder configuration as seen in the literature (N1c in~\cite{Lisa_frequency_planning}), where Eqs.~(\ref{eq.xi}) and (\ref{eq.xip}) are the six one-way ranging signals needed to synthesize Michelson-like interferometers.  Without loss of generality, we consider the laser on MOSA$_{1}$ as the primary laser. The laser in MOSA$_{1'}$ is frequency offset locked to the MOSA$_{1}$ laser. In turn, the laser on MOSA$_{2'}$ is locked to the incoming beam from MOSA$_{1}$ and MOSA$_{3}$ laser to one from MOSA$_{1'}$. MOSA$_{2}$ and MOSA$_{3'}$ are locked to 2' and 3, respectively. 

\subsection{Time Delay Interferometry 2.0}

To cancel the laser phase noise seen in Eqs.~(\ref{eq.xi}) and (\ref{eq.xip}), one must construct TDI variables using these interferometer signals. For this paper, we do an initial study using the Michelson variables that simulate equal-arm interferometers, essentially canceling laser phase noise when the delays are known perfectly.
    
TDI 1.0 fails to suppress LPN when the relative velocity between the SCs is too large. For this reason, TDI 2.0 combinations are needed as it senses each arm twice, while considering flexing arms~\cite{TDI_2p0, TDI_2p0_2}; the second-generation TDI X is essentially a Michelson interferometer with two reflections of the laser beams, which, when including the transponder scheme described above, is given by
\begin{eqnarray} \label{eq.TDI2}
X &=& (\xi_{1:22'} - \xi_{1}) - (\xi_{1':3'3} - \xi_{1'}) \nonumber \\
&& - (\xi_{1:22'22'3'3} - \xi_{1:3'322'}) + (\xi_{1':3'33'322'} - \xi_{1':22'3'3}).
\end{eqnarray}
Moreover, one can always obtain the other basic TDI variables $Y$ and $Z$ through cyclic permutations of the indices; for $Y$: $1\rightarrow 2$, $2\rightarrow 3$ and $3\rightarrow 1$ and similarly for $Z$ starting in $Y$. Now we are ready to start building the TTL model within TDI using Eqs.~(\ref{eq.xi}) and (\ref{eq.xip}). We will then apply a Fisher information treatment after moving to the Fourier domain, where we consider the TTL coupling\footnote{One thing to note is that the following equations consider equal delays for the sake of brevity; the results we present use the full model with different delays, as seen in App.~\ref{app.TDI2p0}. It can be shown that the effect of different delays from the constellation flexing over the mission lifetime has a small impact on the results ($\leq 5\%$ change)} to be the {\it signal} in addition to the rest of the noise terms, i.e., $X = X^{\rm TTL} + {\rm noise}$ 
\begin{eqnarray}
\widetilde{X}^{\rm TTL}&=&\widetilde{H}\times[ \nonumber \label{eq.X.ttl} \\
&& \quad +\widetilde{\mathbf \Psi}_{1}({\bf k}_{1}^{\rm RX}+e^{-i\omega2\tau}{\bf k}_{1}^{\rm TX})     
     -\widetilde{\mathbf \Psi}_{1'}({\bf k}_{1'}^{\rm RX}+e^{-i\omega2\tau}{\bf k}_{1'}^{\rm TX}) \nonumber \\
    && \quad +\widetilde{\mathbf \Psi}_{2'}e^{-i\omega\tau}({\bf k}_{2'}^{\rm RX}+{\bf k}_{2'}^{\rm  TX}) 
    -\widetilde{\mathbf \Psi}_{3}e^{-i\omega\tau}({\bf k}_{3}^{\rm RX}+{\bf k}_{3}^{\rm TX})] 
\label{eq.TDIX_f}
\end{eqnarray}
where
\begin{equation}
    \widetilde{H}=-1+e^{-i\omega2\tau}+e^{-i\omega4\tau}-e^{-i\omega6\tau}
\end{equation}
is the associated transfer function for TDI X, $\tau$ is the delay along the arms, $\omega=2\pi f$ is the angular Fourier frequency, and $\widetilde{\mathbf \Psi}_{i}$ is the Fourier transform of the DWS signal associated with $\mathrm{MOSA}_{i}$. Notice that the coefficients associated with the TX and RX beams for the DWS signals from $\mathrm{MOSA}_{2',3}$ are grouped, meaning one cannot find each coefficient individually using only the TDI X combination. $\widetilde{X}^{\rm TTL}$ (and $\widetilde{Y}^{\rm TTL}$, $\widetilde{Z}^{\rm TTL}$) is the quantity we will analyze in our FIM calculations to identify how well we can estimate each TTL coefficient. Notice that we also make the approximation of time-independent coefficients {\bf k} since expected drifts with the order of $20$\,$\mu$m/rad/day lead to only minor TTL noise.

\subsection{Differential Wavefront Signals} \label{sec.DWS}

We now move on to define the DWS signals themselves and then build up the noise profile for the TDI variable $X$. We are only interested in the relative orientation between the MOSA and the incoming beam in the directions of $\phi$ and $\eta$; $\theta$ (roll ---see Fig.~\ref{fig.LISA_full}) does not play a role due to cylindrical symmetry. These relative motions are determined by the science interferometer DWS signals $\mathbf{\Psi} = (\phi, \eta)$ themselves through the Drag-Free Attitude Control System (DFACS); we do not explicitly consider the control-loops involved in the DFACS, but just assume there is some residual jitter after being suppressed by it. The SC orientation with respect to the inertial frame, or more specifically, with respect to the incoming beam from the far spacecraft will be measured by the DWS, which for yaw is
\begin{eqnarray}
\widetilde{\phi}_{i,i'} = \mathcal{\widetilde{\phi}}_{{\rm SC}_i} + \mathcal{\widetilde{\phi}}_{{\rm M}_{i,i'}} + \mathcal{\widetilde{\phi}}_{\rm n},
\end{eqnarray}
where $\mathcal{\widetilde{\phi}}_{{\rm SC}_i}$ is the SC jitter in yaw, $\mathcal{\widetilde{\phi}}_{{\rm M}_{i,i'}}$ is the MOSA jitter with respect to the SC inertial frame, and $\mathcal{\widetilde{\phi}}_{\rm n}$ is the readout noise associated with a DWS measurement. We assume the readout noise is not imprinted onto the actual angles since the DFACS is gain limited. For $\eta$ (pitch)
\begin{eqnarray}
\mathcal{\widetilde{\eta}}_{i} & = \mathcal{\widetilde{\eta}}_{{\rm SC}_i}\cos{30^{\mathrm o}} - \widetilde{\mathcal{\theta}}_{{\rm SC}_i}\sin{30^{\mathrm o}} + \mathcal{\widetilde{\eta}}_{{\rm M}_i} \label{eq.eta1} + \mathcal{\widetilde{\eta}}_{\rm n}
\\
\mathcal{\widetilde{\eta}}_{i'} & = \mathcal{\widetilde{\eta}}_{{\rm SC}_i}\cos{30^{\mathrm o}} + \widetilde{\mathcal{\theta}}_{{\rm SC}_i}\sin{30^{\mathrm o}} + \mathcal{\widetilde{\eta}}_{{\rm M}_{i'}} + \mathcal{\widetilde{\eta}}_{\rm n}
\label{eq.eta2}
\end{eqnarray}
since the pitch and roll of the SC are coupled with each other asymmetrically ---see Fig.~\ref{fig.LISA_full}. For the calculations, we use two cases: (i) the current LISA requirements~\cite{lisa_reqs} as case A, and (ii) the predicted jitter from LISASim (a time-domain LISA simulator described in~\cite{lisaSIM}) as case B. The signals are defined in Appendix~\ref{app.jitt.sign} and shown in Fig.~\ref{fig.jitter}.

%
\begin{figure}[h!]
\begin{center}
    \includegraphics[width=\linewidth]{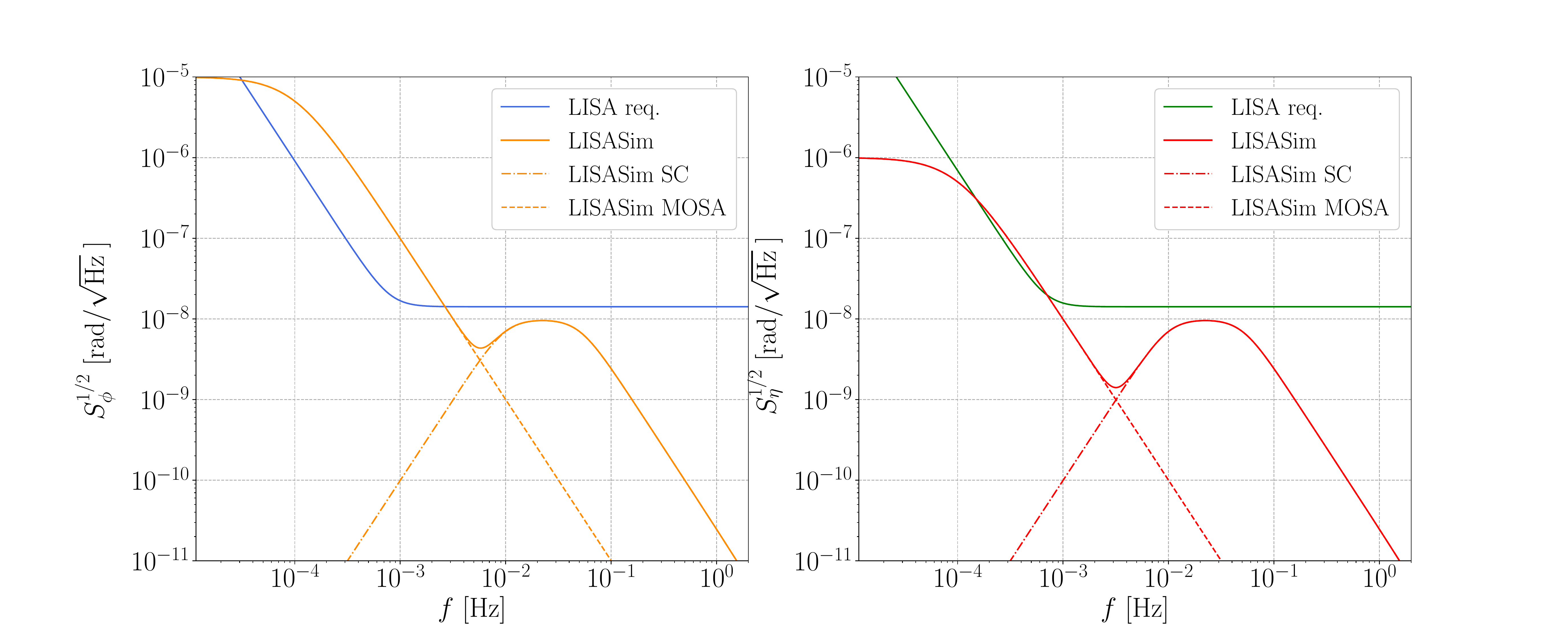}
\end{center}
\caption{Total angular jitter (left: yaw, right: pitch) profiles considered for the LISA requirement (case A) and total, SC, and MOSA jitter profiles from LISASim (case B). We do not display the SC and MOSA curves of case A as they are all similar.}
\label{fig.jitter}
\end{figure}
\subsection{Noise sources}
\label{sec:noise}

\begin{figure}[h!]
\begin{center}
    \includegraphics[width=0.7\linewidth]{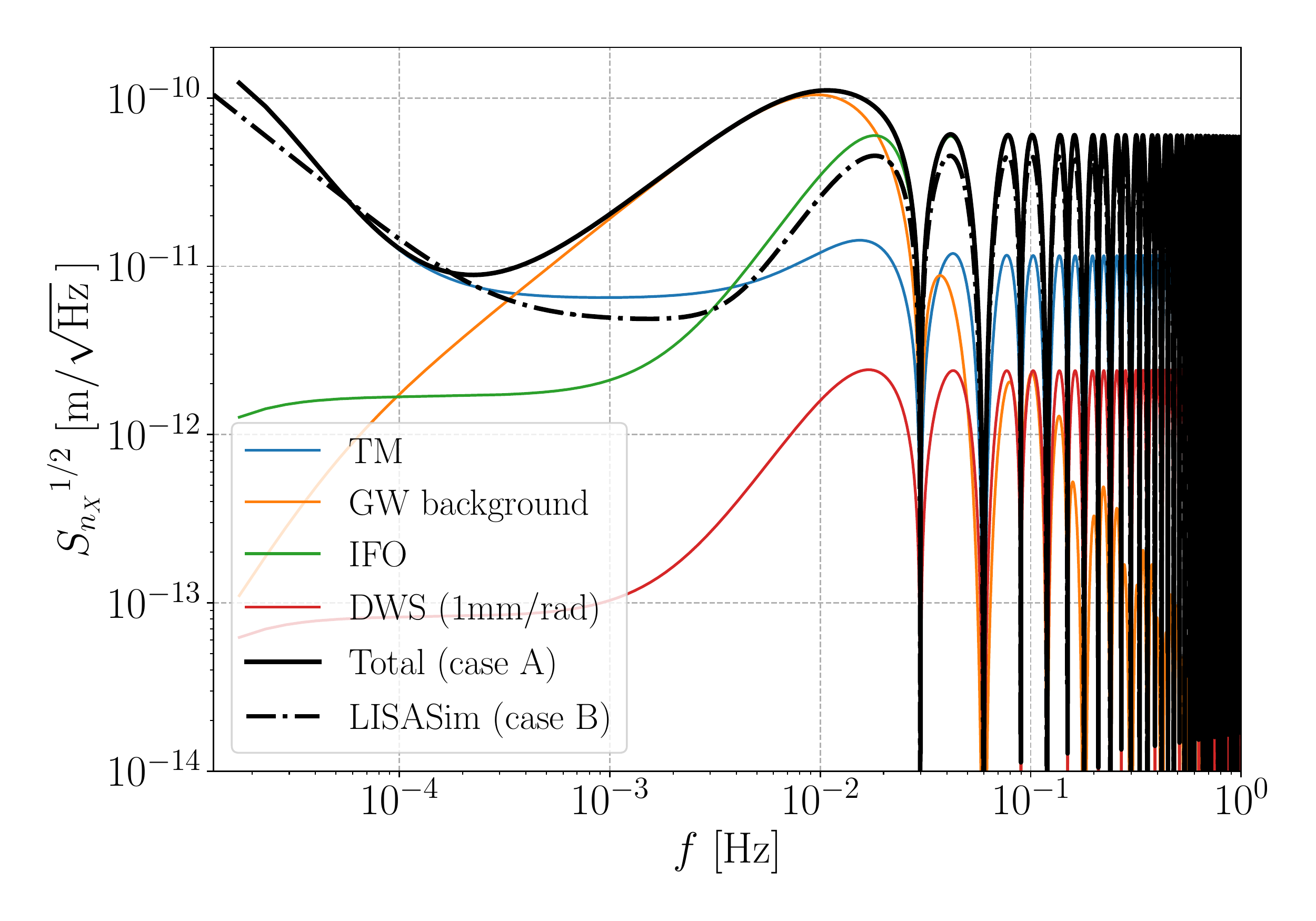}
\end{center}
\caption{Noise contributions for case A after going through second-generation TDI X. TM: test-mass acceleration; IFO: IFO noise; DWS: DWS readout noise. Case B's noise curve is given by LISASim total. Delays used: $\tau_{2,2'} = 8.29 \;\rm{s}; \: \tau_{3,3'} = 8.40 \;\rm{s}$.} 
\label{fig.noise}
\end{figure}

Accurate estimation of TTL coefficients require sufficient signal-to-noise Ratio (SNR) in the length noise. Subsequent subtraction of TTL noise also requires adequate SNR in the DWS readout. Assuming LPN is already removed with the help of TDI, we consider the secondary noise sources to be from residual TM acceleration, interferometric noise (including the science, the local and the reference interferometers, and the back-link fiber), GW background, and DWS readout noise. The effect of DWS noise is two-fold as it plays a role in both the estimation of TTL coefficients and the subtraction of TTL noise, which will be seen later. Here we list the noise sources associated with case A for equal delays, and leave the full expressions in Appendix~\ref{app.noise}. For case B (LISASim), one can refer to~\cite{lisaSIM} for the individual sources. Assuming none of these sources are correlated with each other, the total noise power spectral density (PSD) is
%
%
\begin{eqnarray}
S_{\rm n} &=& S_{\rm {IFO}} + S_{\rm {TM}} +S_{\rm {DWS}} + S_{\rm {GW}}. \label{eq.sn.all}
\end{eqnarray}
Keep in mind that each term has been propagated through the TDI transfer functions, and is not simply an addition of the LISA noise estimates; each term can be seen in Fig.~\ref{fig.noise}. The main difference in the two cases is that case A considers a GW background, while case B does not; one can ascertain that the noise profiles are very similar apart from this feature. Due to TDI, the interferometric noise transforms to
%
%
\begin{equation}
S_{\rm {IFO}} = 4|\widetilde{H}|^{2}S_{\rm s}
\end{equation}
where 
\begin{equation}
    |\widetilde{H}|^{2}=16\sin^{2}\omega\tau\sin^{2}2\omega\tau,
\end{equation}
and $S_{\rm s}$ is the interferometer noise defined in Eq.~(\ref{eq.intro.2a}). Similarly, the residual TM acceleration motion is transformed by TDI into the form 
\begin{equation}
S_{\rm {TM}} = \frac{4|\widetilde{H}|^{2}}{\omega^{2}}(3 + \cos{2\omega\tau})S_{\delta}
\end{equation}
with $S_{\delta}$ the residual acceleration in a single TM and defined in Eq.~(\ref{eq.TM_noise}).
Since we are simultaneously considering the TTL associated with jitter in both pitch and yaw, we have to consider the total DWS measurement noise $S_{\rm {DWS}} = S_{\rm {\phi}} + S_{\rm {\eta}}$. Strictly speaking, the DWS noise in terms of displacement noise is unknown since the TTL couplings are yet to be estimated, for e.g., one TTL term in yaw $c_1 (\phi_1 + \phi_{n_1})$ would have a DWS noise of $c_1 \phi_{\rm n_1}$. This issue can be temporarily circumvented if we assign a value of $3 \mathrm{\, mm/rad}$ per coefficient. If one assumes all the coefficients to be equal, and all the jitters are correlated, and the same noise level for yaw and pitch, we get
\begin{equation}
    S_{\rm {DWS}}=8k^{2}|\widetilde{H}|^{2}(3 + \cos{2\omega\tau})S_{\rm n_{DWS}}.
\end{equation}
where $S_{\rm n_{DWS}}$ is given in Eq.~(\ref{eq.noise.dws}).
Although DWS length noise does not play much of a role in estimating the TTL coefficients, it can be a limiting factor when one subtracts TTL noise in TDI, as seen in section~\ref{sec:results}. The degree to which it will limit us will be based on the initial values of TTL coefficients themselves, as the DWS noise projection onto TDI will scale by the coefficients. Finally, the gravitational wave background, $S_{\mathrm h}$ is roughly based on the Mock LISA Data Challenges (MLDC) data set~\cite{Babak_2008} that considers massive black holes, extreme mass ratio inspirals, cosmic-string bursts, an isotropic stochastic background, and a large galactic binary population. It appear in the TDI combination as
\begin{equation}\label{eq.Sh}
    S_{\rm GW}=4L^{2}|{\rm sinc(\omega\tau)}|^{2}|\widetilde{H}|^{2}\sin^{2}60^{\rm o}S_{\rm h},
\end{equation}
where $L$ is the LISA's arm length, 2.5\, Gm, and $S_{\rm h}$ is defined in the appendix, Eq.~(\ref{eq.GW.app}).

The signals and noise floors we defined in both cases will dictate the uncertainty associated with recovering each TTL coefficient. With the finite SNR we have in both cases, the Fisher information will yield the associated uncertainty without explicitly finding the values of the coefficients themselves using time-domain simulations. 

\section{Fisher Information of TTL within TDI} \label{sec:FIM.in.TDI}

The Fisher Informaion Matrix (FIM)~\cite{FIM, FIM_2,PhysRevD.77.042001} in our case quantifies the amount of information a TDI variable has encoded in the TTL coefficients vector $\bf k$. It is given by 
\begin{equation}
\label{eq.FIM.int}
F_{i,j} = 4 \int_{\omega_{\rm min}}^{\omega_{\rm max}}\frac{{\rm d}\omega}{2\pi}\frac{1}{S_{\rm n}(\omega)} {\rm  Re}{\bigg\{\bigg(\frac{\partial \widetilde{X}^{\rm TTL}}{\partial k_i^{\rm RX,TX}}\bigg)^*  \bigg(\frac{\partial \widetilde{X}^{\rm TTL}}{\partial k_{j}^{\rm RX,TX}}\bigg) \bigg\}},
\end{equation}
where $\omega_{\rm min} = 2\pi/{T}$ accounts for the integration time $T$, $\omega_{\rm max} = 2\pi f_{\rm s}/2$ considers the sampling frequency $f_{\rm s}$, $S_{\rm n}$ is the one-sided PSD of the noise defined in Eq.~(\ref{eq.sn.all}), $\widetilde{X}^{\rm TTL}$ is the TTL model presented in Eq.~(\ref{eq.TDIX_f}), and $k^{\rm RX,TX}$ are the TTL coefficients we are interested in. The covariance matrix (inverse of the FIM) indicates how well we can recover individual coefficients but is also used to compute the TTL residual. Herein lies the crux of this paper. We will use the information from the previous sections that built the TTL model to constrain the TTL coefficients using the FIM and estimate the residual TTL contribution to the overall LISA noise curve. The Cram\'{e}r-Rao bound predicts the lower limit of the errors on the TTL coefficients without actually obtaining the posteriors on the true values
\begin{equation}
{\sigma^2}({\bf k}) \geq {\rm diag}(\mathbf{F}^{-1}).
\end{equation}

Singular matrices will also inform us that the DWS signals associated with an individual TTL coefficient are too similar to another, and therefore the two (or more) coefficients cannot be estimated independently. The only possible recourse at that point is to reduce the dimensionality of the FIM by combining the coefficients with similar DWS signals~\cite{PhysRevD.77.042001}. Notice this still allows to subtract TTL effects but prevents from knowing all the TTL coefficients individually. However, all the coefficients can, in principle, be found independently when considering the three TDI combinations $X$, $Y$, and $Z$.

The FIM for the current formulation of both yaw, $\phi$, and pitch, $\eta$ is given in Eq.~(\ref{eq.FIM}). Note that this is the FIM for the $X$ combination and includes $\phi$ and $\eta$ coefficients, i.e, ${\mathbf k}$=\{$c$,$d$\}. The matrix is sized 12$\times$12 due to considering the coefficients that appear in the $X$ combination after grouping coefficients that are fully correlated ---see Eq.~(\ref{eq.X.ttl}), otherwise, the matrix would have been 16$\times$16 (4 MOSAs $\times$ 2 coefficients $\times$ 2 degrees of freedom). 

\begin{equation}{\label{eq.FIM}}
{\mathbf F^{\widetilde{X}}} = \left[\begin{array}{ c | c }
    {\mathbf F^{\widetilde{X}}_{\phi}} & 0 \\
    \hline
    0 & {\mathbf F^{\widetilde{X}}_{\eta}}
  \end{array}\right]
\end{equation}
where the upper right and lower left quadrants are zero because we assume that yaw and pitch jitter are orthogonal to each other, which holds if the DWS read-out has no cross-talk. 
\begin{equation}{\label{eq.FIM_phi}} {\mathbf F^{\widetilde{X}}_{\phi}} = 
\begin{blockarray}{cccccc|cccccrl}
{\scriptstyle c_{1}^{\rm RX}} & {\scriptstyle c_{1}^{\rm TX}} & {\scriptstyle c_{1'}^{\rm RX}} 
& {\scriptstyle c_{1'}^{\rm TX}} & {\scriptstyle c_{2'}^{\rm RX}+c_{2'}^{\rm TX}} & {\scriptstyle c_{3}^{\rm RX}+c_{3}^{\rm TX}} \\
\cline{1-13}
\begin{block}{[cccccc]|cccccrl}
  F_{1,1} & F_{1,2} & F_{1,3} & F_{1,4} & 0 & 0 && {\scriptstyle c_{1}^{\rm RX}}\\
    F_{1,2} & F_{1,1} & F_{2,3} & F_{2,4} & 0 & 0 && {\scriptstyle c_{1}^{\rm TX}}\\
    F_{1,3} & F_{2,3} & F_{3,3} & F_{3,4} & 0 & 0 && {\scriptstyle c_{1'}^{\rm RX}}\\
    F_{1,4} & F_{2,4} & F_{3,4} & F_{3,3} & 0 & 0 && {\scriptstyle c_{1'}^{\rm TX}}\\
    0  & 0 & 0 & 0 & F_{5,5} & 0 && {\scriptstyle c_{2'}^{\rm RX}}+{\scriptstyle c_{2'}^{\rm TX}}\\
    0  & 0 & 0 & 0 & 0 & F_{6,6} && {\scriptstyle c_{3}^{\rm RX}}+{\scriptstyle c_{3}^{\rm RX}}\\
\end{block}
\end{blockarray}
\end{equation}
\begin{equation}{\label{eq.FIM_eta}} {\mathbf F^{\widetilde{X}}_{\eta}} = 
\begin{blockarray}{cccccc|ccccccrl}
{\scriptstyle d_{1}^{\rm RX}} & {\scriptstyle d_{1}^{\rm TX}} & {\scriptstyle d_{1'}^{\rm RX}} 
& {\scriptstyle d_{1'}^{\rm TX}} & {\scriptstyle d_{2'}^{\rm RX}+d_{2'}^{\rm TX}} & {\scriptstyle d_{3}^{\rm RX}+d_{3}^{\rm TX}}\\
\cline{1-13}
\begin{block}{[cccccc]|ccccccrl}
    F_{7,7} & F_{7,8} & F_{7,9} & F_{7,10} & 0 & 0 && {\scriptstyle d_{1}^{\rm RX}}\\
    F_{7,8} & F_{7,7} & F_{8,9} & F_{8,10} & 0 & 0 &&{\scriptstyle d_{1}^{\rm TX}}\\
    F_{7,9} & F_{8,9} & F_{9,9} & F_{9,10} & 0 & 0 &&{\scriptstyle d_{1'}^{\rm RX}}\\
    F_{7,10} & F_{8,10} & F_{9,10} & F_{9,9} & 0 & 0 &&{\scriptstyle d_{1'}^{\rm TX}}\\
    0 & 0 & 0 & 0 & F_{11,11} & 0&&{\scriptstyle d_{2'}^{\rm RX}}+{\scriptstyle d_{2'}^{\rm RX}}\\
    0 & 0 & 0 & 0 & 0 & F_{12,12}& & {\scriptstyle d_{3}^{\rm RX}}+{\scriptstyle d_{3}^{\rm RX}}\\
\end{block}
\end{blockarray}
\end{equation}
where, for example
\begin{eqnarray}
    F_{1,1}&=&\int \frac{{\rm d}\omega}{2\pi}\frac{|\widetilde{\phi}_{\rm SC_{1}}|^{2}+|\widetilde{\phi}_{\rm M_{1}}|^{2}}{S_{\rm n}}\\
    F_{1,2}&=&\int \frac{{\rm d}\omega}{2\pi}\frac{|\widetilde{\phi}_{\rm SC_{1}}|^{2}+|\widetilde{\phi}_{\rm M_{1}}|^{2}}{S_{\rm n}}\cos{\omega2\tau} \\
    F_{1,3}&=&-\int \frac{{\rm d}\omega}{2\pi}\frac{|\widetilde{\phi}_{\rm SC_{1}}|^{2}}{S_{\rm n}} \\
    F_{7,7}&=&\int \frac{{\rm d}\omega}{2\pi}\frac{|\widetilde{\eta}_{\rm SC_{1}}|^{2}\cos^{2}{30^{\rm o}}+|\widetilde{\theta}_{\rm SC_{1}}|^{2}\sin^{2}{30^{\rm o}}+|\widetilde{\eta}_{\rm M_{1}}|^{2}}{S_{\rm n}}
\label{eq.some_F_terms}
\end{eqnarray}
considers TTL coefficients $c_{1}^{\rm RX}, \: c_{1}^{\rm TX}, \: d_{1}^{\rm RX}, \: d_{1}^{\rm TX}$. $F_{1,1}$ computes the curvature in the likelihood function purely due to $c_{1}^{\rm RX}$. Physically, $c_{1}^{\rm RX}$ refers to the TTL coupling appearing on $\rm MOSA_1$ due to local jitter in yaw. On the other hand, $c_{1}^{\rm TX}$ is also measured on $\rm MOSA_1$ due to the transponder scheme onboard SC$_{2}$ that maintains the phase information of TTL due to jitter on SC$_{1}$ in the transmit beam, and sends it back to SC$_{1}$, resulting in a delay of $2\tau$, which shows up in $F_{1,2}$ and indicates the cross-correlation between $c_{1}^{\rm RX}$ and $c_{1}^{\rm TX}$. One can also notice that the cross-correlation in $F_{1,3}$ occurs due to SC jitter being common to $\rm MOSA_1$ and $\rm MOSA_{1'}$ in yaw. Moreover, since $S_{\phi} = |\widetilde{\phi}|^2/T$, $F_{i,j}$ will then depend linearly on $T$, meaning the uncertainties on the TTL coefficients will scale as $1/\sqrt{T}$.

We assume that the jitter on board one SC is not correlated to another, meaning there are no off-diagonal terms for $c_{2'}^{\rm RX} + c_{2'}^{\rm TX}$ and $c_{3}^{\rm RX} + c_{3}^{\rm TX}$, which in turn means that, in principle, these grouped coefficients can be recovered with lower uncertainties in TDI X. One can also repeat the exercise with the other TDI variables and apply the same reasoning for the pitch degree of freedom as well. 
%
%
    
\section{Effects of angular jitter spectrums on TTL calibration \label{sec:results}}
Here, we present the effects of considering the two cases introduced in Sec.~\ref{sec.DWS} (Fig.~\ref{fig.jitter}), with their similar noise profiles shown in Fig.~\ref{fig.noise}; case A considers a GW background while case B does not, however setting $S_{\mathrm h} = 0$ has a minimal effect. To cross-check the Fisher information results, we perform time-domain simulations as well: in case A, we inject ${\bf k}_{\text{inj}}$ from a random uniform distribution  with amplitude of $\pm 5 \: {\rm mm/rad}$to simulate TDI data with appropriately colored noise sources and delays as in Eq.~(\ref{eq.TDI2}); case B uses data simulated by LISASim also from the same distribution. The injections are then recovered using a weighted least-squares (WLS) in the frequency domain. A Monte-Carlo approach is used to build a distribution where the standard deviations can be compared with the Cram\'{e}r-Rao inequality, which gives the theoretical lower-bound on $\bf k$ that any time-domain simulation result would hope to approach.

Table~\ref{table:results} shows the errors from the FIM in columns 1 and 3 for cases A and B respectively, and columns 2 and 4 show the results from the respective time-domain simulations used for cross-checking the FIM results. In both cases, the FIMs are calculated assuming a sampling rate of 4\,Hz and a measurement duration of approximately one day; this sets the integral limits in Eq.~(\ref{eq.FIM.int}) from $\sim$11\,$\mu$Hz to 2\,Hz. Table~\ref{table:results} shows agreement within $\simeq15\%$ for all coefficients with the time-domain LS simulations. Case A outperforms case B by having standard deviations that are about $50 \times$ smaller for the individual coefficients, and about $5 \times$ smaller for the grouped coefficients. However, the physical reason for this has to be taken into consideration: case A has jitter that is not very realistic since it does not capture the natural roll-off of jitter expected at higher frequencies due to SC inertia. Therefore, one has to be careful when using the LISA requirements in their jitter models for TTL calibration as it does not capture this roll-off at higher frequencies.

There are other subtleties to consider that do not make the FIM and LS a perfect apples-to-apples comparison: due to the linear model of TTL coupling we use, the Fisher information does not depend on the value of the actual coefficients. However, in the time-domain implementation, the appearance of DWS noise in the length noise will be scaled by the injected TTL coefficients. Further work is required to consider this subtlety within the Fisher information formalism. Even so, the columns do agree well enough to show that these are not significant issues. Note that these values also hold for the TDI $Y$ and $Z$ combinations as well by cyclic permutation of the indices.

\begin{table}[h!]
\begin{center}
\begin{tabular}{lcc}
\toprule
 Coeffs. & FIM Case A (LS) & FIM Case B (LS) \\
in TDI $X$ & [$\mu$m/rad] & [$\mu$m/rad] \\ \midrule

$c_{1}^{\rm RX}$  & 3.23 (2.97) & 206 (209)\\
$c_{1}^{\rm TX}$  & 3.23 (2.91) & 200 (205)\\
$c_{1'}^{\rm RX}$  & 3.23 (2.88) & 205 (208)\\
$c_{1'}^{\rm TX}$ & 3.23 (2.93) & 200 (205)\\
$c_{2'}^{\rm RX} + c_{2'}^{\rm TX}$  & 2.91 (2.59) & 25.4 (24)\\
$c_{3}^{\rm RX} + c_{3}^{\rm TX}$  & 2.90 (2.53) & 25.4  (29)\\ \midrule
$d_{1}^{\rm RX}$  & 2.98 (2.74) & 245 (262)\\
$d_{1}^{\rm TX}$  & 2.98 (2.82) & 212 (260)\\
$d_{1'}^{\rm RX}$  & 2.98 (2.71) & 245 (258)\\
$d_{1'}^{\rm TX}$  & 2.98 (2.79) & 213 (262)\\
$d_{2'}^{\rm RX} + d_{2'}^{\rm TX}$  & 2.87 (2.65) & 29.9 (31)\\
$d_{3}^{\rm RX} + d_{3}^{\rm TX}$  & 2.86 (2.68) & 29.9 (27) 
 \\\bottomrule
\end{tabular}
 \caption{1-sigma standard deviations on ${\bf k} = {c, d}$ obtained from two noise and jitter profiles using the FIM approach and the corresponding time-domain simulations. Case A: Using LISA requirement angular jitter for $T$ = 86400 s; case B: Using LISASim angular jitter for $T$ = 76400 s. LS: parameters estimated from Monte Carlo time-domain simulations using weighted least-squares. These numbers were calculated using TDI $X$, but also holds for $Y$ and $Z$}
 \label{table:results}
\end{center}
\end{table}

\begin{figure}[h!]
\begin{center}
    \includegraphics[width=0.7\linewidth]{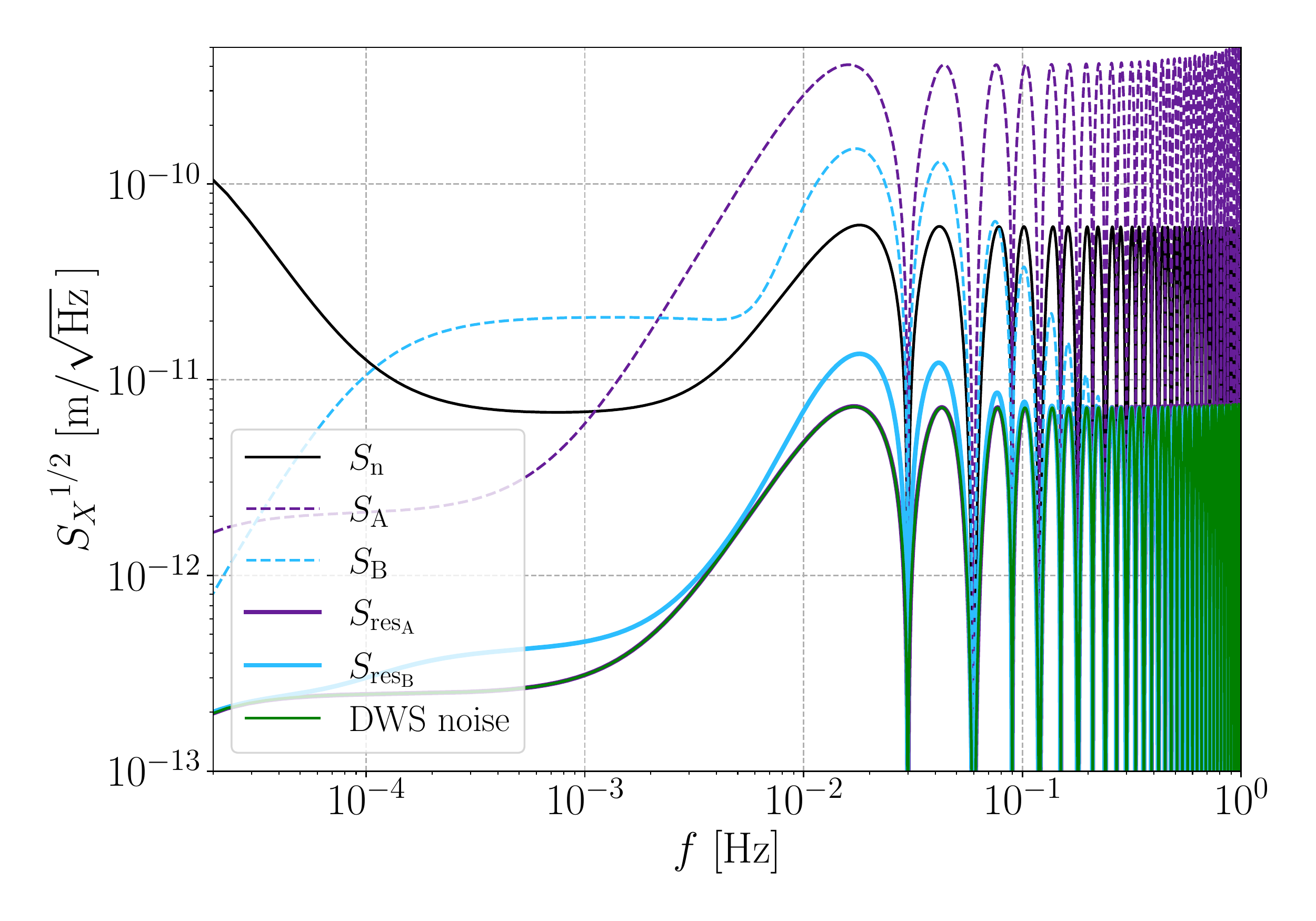}
\end{center}\caption{TDI X residual computed from the error covariance matrix $\Sigma$. $S_{\rm n}$: All noise sources as seen in Fig.~\ref{fig.noise} for case A, excluding  the gravitational background; $S_{\rm A}, \: S_{\rm B}$: Noise contributions from TTL assuming $\bf k=\rm 3 \:mm/rad$; $ S_{\rm residual}$: Expected residual ${S_{\rm res_A}}$ (${S_{\rm res_B}}$) for case A (case B) using expected jitter and an integration time of $T \sim 1$ day as computed in Eq.~(\ref{eq.res}); case A has a roughly $6 \times$ lower residual. DWS: readout noise assuming $\bf k=\rm 3 \:mm/rad$.  \label{fig.residual}}
\end{figure}

Though finding individual components of TTL is useful in pinpointing locations of potentially problematic sources, we are more interested in finding the residual TTL noise after subtraction in post-processing. The error covariance matrix ${\mathbf \Sigma} = {\mathbf F}^{-1}$ allows us to find the theoretical residual (see App.~\ref{app.residual} for the full equation using no approximations of the delays)
\begin{eqnarray}\label{eq.res}
S_{\rm res, \phi} &=& (\tilde{X}-\hat{\tilde{X}})_{\phi}(\tilde{X}-\hat{\tilde{X}})_{\phi}^{*}  \nonumber \\ 
&=& \{ S_{\phi_{\rm SC_{1}}} [\sigma_{11}^2 + \sigma_{22}^2 + \sigma_{33}^2 +\sigma_{44}^2 - 2(\sigma_{13}+\sigma_{24})
+2(\sigma_{12}+\sigma_{34}-\sigma_{14}-\sigma_{23})\cos{2\omega\tau}] \nonumber \\
&&+ S_{\phi_{\rm M_{1}}} [\sigma_{11}^2 + \sigma_{22}^2 + 2\sigma_{12}\cos{2\omega\tau}] \nonumber  \\
&&+ S_{\phi_{\rm M_{1'}}} [ \sigma_{33}^2 + \sigma_{44}^2
+ 2\sigma_{34}\cos{2\omega\tau}] \nonumber \\
&&+ (S_{\phi_{\rm SC_{2}}}+S_{\phi_{\rm M_{2'}}}) \sigma_{55}^2 + (S_{\phi_{\rm SC_{3}}}+S_{\phi_{\rm M_{3}}}) \sigma_{66}^2  \nonumber \\
&&+S_{\rm {DWS\,\phi}} 
\end{eqnarray}
where $\sigma_{ij}^{2}$ are the elements of ${\mathbf \Sigma}$, $S_{\phi_i}$ are the respective angular jitters, and $S_{\rm {DWS}_{\phi}}=4k^{2}|\widetilde{H}|^{2}(3 + \cos{2\omega\tau})S_{\rm n_{DWS}}$ is the DWS readout noise projection onto TDI X scaled by TTL coupling. A similar treatment can be performed to obtain the total residual including pitch ($\eta$), giving $S_{\mathrm{res}} = S_{\mathrm{res},\phi} + S_{\mathrm{res}, \eta}$ as seen in Fig.~\ref{fig.residual}. Note t hat the TTL uncertainty contribution to the TTL noise residual is independent of the coefficients themselves: the coefficients appear in the residual through the DWS readout noise $S_{\rm {DWS}}$. In case A, we are limited by the DWS measurement noise projection onto TDI X, which is given assuming $\bf k = \rm 3 \:mm/rad$. Case B is limited by the TTL coefficients' uncertainties that are roughly $70\times$ greater, occurring due to the lower SC jitter at high frequencies, along with the DWS noise projection.
If one assumes all the TTL coefficients $\bf k = \rm 3 \:mm/rad$, we find suppression of about $80\times$ ($11\times$) in case A (case B) predicted by the covariance matrix, such that the TTL residual in case A is around $8 \: \rm pm/\sqrt{Hz}$ in one day of integration time. However, a calibration of the TTL coefficients over $T = 76400 \: \rm s$ for case B reaches a maximum residual TTL noise of $13.6 \: \rm pm/\sqrt{Hz}$ at $20 \rm \: mHz$ falling off on either side in the frequency-domain. Moreover, both residuals are below the projected noise floor for the case that all the coefficients are equal $\bf k=\rm 3 \:mm/rad$. At higher TTL coefficient values, DWS readout noise will dominate, erasing the effect of the two angular jitter models in the TTL noise residual. In the event that TTL coefficients are $\gtrsim{20 \: \rm mm/rad}$, one might have to implement an in-orbit alignment procedure to ensure DWS readout noise does not limit the interferometer performance when trying to subtract TTL noise.

Since the length noise curves in both cases are very similar, as seen in Fig.~\ref{fig.noise} (apart from the GW background in case A which has a small effect), the differences in performance must be purely due to the differences in angular jitter, in particular at higher frequencies ---see Fig.~\ref{fig.jitter}. The angular noise power at higher frequencies contributes the most SNR in the Fisher information, giving 70$\times$ lower uncertainties in individual TTL coefficients, and a 1.7$\times$ lower noise residual. However, as stated earlier, case A is a jitter requirement while case B is a model of the expected jitter based on a specific and simple DFACS model. 

\section{Discussion \label{sec:final}}

We have shown that the Fisher information analysis allows us to predict how well one can suppress noise due to TTL coupling after TDI removes laser phase noise. The Cram\'{e}r-Rao bound gives the lower bound we have in estimating TTL coefficients, thereby telling us the accuracy to which we can theoretically resolve them. We find our results agree well with Monte-Carlo time-domain simulations from our own code, and with LISASim. The Fourier domain approach allows one to quickly make calculations of how well TTL can be suppressed based on the spectrum of angular jitter. Our approach allows us to modify noise profiles, use various angular jitter spectra, or find the minimum integration times required to be below the noise floor.

We considered two main cases in this work: case A utilizes the LISA requirement angular jitter, while case B considers a more realistic angular jitter from LISASim. The two cases have very similar noise floors, while the former also considers a GW background, which does not have much of an effect on our results. We find the unrealistic levels of jitter above $20 \: \rm mHz$ in case A leads to an unrealistic TTL calibration residual below $2 \: \rm pm/\sqrt{Hz}$ for all frequencies in one day of measurement time. On the other hand, the residual in case B is approximately below $12 \: \rm pm/\sqrt{Hz}$ across all frequency bins. Both cases are at least $5\times$ below the length noise floor, even though the uncertainties associated with individual TTL coefficients for case B were high, as seen in Table~\ref{table:results}. We also find that care has to be taken when the system has a larger magnitude of TTL coefficients, as this will translate to residuals approaching the length noise floor.

We also find that using the TDI $X$, $Y$, $Z$ variables, one can find the values of every TTL coefficient independently. This is important to identify the TTL coefficients across the six MOSAs and re-align them. 
If considering only TDI X, we see that we cannot distinguish between the RX and TX coefficients from MOSA$_{2'}$ and MOSA$_{3}$ as seen in Eq.~(\ref{eq.TDIX_f}). 

There are other questions we can answer with the Fisher information formalism: firstly, do other TDI forms give us more information? As an initial step, we used the Michelson combinations, but we can also consider the Sagnac, orthogonal, or other TDI combinations presented in~\cite{Muratore_2022}. Secondly, we use given noise models, whereas there could be several unknowns in a space mission. In future work, we will consider the case of having no prior knowledge of the noise profile, and find it e.g. with an iterative reweighted least-squares or log-likelihood as done in~\cite{S_Vitale_2014}. Thirdly, by finding TTL coefficients in regular intervals, one can notice the trend at which TTL changes over time. Any deviation from a nominal trend could point to possible issues such as a misalignment in the OB, unwanted temperature gradients, or other effects. Finally, the final performance of the DFACS system is still under investigation such that the residual spacecraft jitter that determines the required integration time to calibrate TTL sufficiently well is not well known. It might even be necessary to inject spacecraft jitter~\cite{nofrarias_2016}, as done for example in~\cite{Houba_2022} to calibrate TTL swiftly.  

\section*{Acknowledgments}

We thank Alexander Weaver, Sourath Ghosh, and Paul Edwards for useful discussions. We also thank Martin Hewitson, Roberta Giusteri, and Sarah Paczkowski for providing us with TDI simulated data from LISASim. This work was  supported by the National Aeronautics and Space Administration under grants 80NSSC20K0126, 80NSSC22K0675 and 80NSSC19K0324. 


\appendix
\section{Interferometer signals}

These are the science interferometer signals after using the method used in~\cite{TDI_otto} to remove OB motion, with a phase lock loop as defined in the laser frequency transponder scheme~\cite{Lisa_frequency_planning}.
\begin{eqnarray} \label{eq.xi_signals1}
    \xi_{1} &=& p_{1:3'3} - p_{1} + n_{2':3} + n_{1} \nonumber \\
    &&+ {\bf e}_{3}(\delta_{1}+\delta_{1:3'3}) + 2{\bf e}_{3'}\delta_{2':3}
    \nonumber  \\
    &&+ h_1 + h_{2':3} \nonumber  \\
    &&+ \Delta\theta_{2':3} \\
    &&+ {\mathbf k}^{\rm TX}_{1}{\mathbf \Psi}_{1:3'3} +  {\mathbf k}^{\rm RX}_{1}{\mathbf \Psi}_{1} +  
    ({\mathbf k}^{\rm RX}_{2'} + {\mathbf k}^{\rm TX}_{2'}){\mathbf \Psi}_{2':3}   \\
    \xi_{1'} &=& p_{1:22'} - p_{1} +  n_{3:2'} + n_{1'}  \nonumber \\
    &&+ {\bf e}_{2'}(\delta_{1'}+\delta_{1':22'}) + 2{\bf e}_{2}\delta_{3:2'}
    \nonumber \\
    &&+ h_{1'} + h_{3:2'} \nonumber  \\
    &&+ \Delta\theta_{3:2'} \\
    &&+ {\mathbf k}^{\rm TX}_{1'}{\mathbf \Psi}_{1':22'} + {\mathbf k}^{\rm RX}_{1'}{\mathbf \Psi}_{1'} + ({\mathbf k}^{\rm RX}_{3} + {\mathbf k}^{\rm TX}_{3}){\mathbf \Psi}_{3:2'} \\
    \xi_{2}&=&p_{1:21}-p_{1:3'} + n_{3:1}  + n_{2} - n_{2'}  \nonumber \\
    &&+\mathbf{e}_{1}\delta_{2}+\mathbf{e}_{1'}\delta_{3':1}+
    \mathbf{e}_{2}\delta_{3:1}+\mathbf{e}_{2'}\delta_{1':21}
    \nonumber \\
    &&-\mathbf{e}_{3}\delta_{1:3'}-\mathbf{e}_{3'}\delta_{2'}
    \nonumber \\
    &&+ h_2 - h_{2'} + h_{3:1} \nonumber  \\
    &&+{\bf k}_{3}^{\rm RX}{\mathbf \Psi}_{3:1}+{\bf k}_{1'}^{\rm TX}{\mathbf \Psi}_{1':21}-{\bf k}_{2'}^{\rm RX}{\mathbf \Psi}_{2'}-
    {\bf k}_{1}^{\rm TX}{\mathbf \Psi}_{1:3'} \nonumber \\
    &&+{\bf k}_{2}^{\rm RX}{\mathbf \Psi}_{2}+{\bf k}_{3'}^{\rm TX}{\mathbf \Psi}_{3':1} \\
    \xi_{2'} &=& \Delta\theta_{2'} \\
    \xi_{3} &=& \Delta\theta_{3} \\
    \xi_{3'}&=&p_{1:3'1'}-p_{1:2} + n_{2':1'} + n_{3'} - n_{3}   \nonumber \\
    &&+\mathbf{e}_{1}\delta_{2:1'}+\mathbf{e}_{1'}\delta_{3'}-
    \mathbf{e}_{2}\delta_{3}-\mathbf{e}_{2'}\delta_{1':2}
    \nonumber \\
    &&+\mathbf{e}_{3}\delta_{1:3'1'}+\mathbf{e}_{3'}\delta_{2':1'}
    \nonumber \\
    &&+ h_{3'} - h_{3} + h_{2':1'} \nonumber  \\
    &&+{\bf k}_{2'}^{\rm RX}{\mathbf \Psi}_{2':1'}+{\bf k}_{1}^{\rm TX}{\mathbf \Psi}_{1:3'1'}-{\bf k}_{3}^{\rm RX}{\mathbf \Psi}_{3}-
    {\bf k}_{1'}^{\rm TX}{\mathbf \Psi}_{1':2} \nonumber \\
    &&+{\bf k}_{3'}^{\rm RX}{\mathbf \Psi}_{3'}+{\bf k}_{2}^{\rm TX}{\mathbf \Psi}_{2:1'}.\label{eq.xi_signals2}
\end{eqnarray}
where $p_i$ is the laser phase noise, $n_i$ is the interferometer noise, ${\bf e}\delta$ is the projection of the test-mass motion along the sensitive axis, $h_i$ is a gravitational signal, and $\Delta\theta$ is the phase error induced from an imperfect phase-lock loop (PLL) in the transponder SC. ${\mathbf k} = (c, d)$ are the TTL coefficients associated with the yaw and pitch degrees of freedom respectively. These angular degrees of freedom are measured by the DWS signals $\mathbf{\Psi} = (\phi, \eta)$.
%

\section{Noise and Jitter Spectra}
\subsection{Noise \label{app.noise}}
Here we show the equations for the curves given in Fig.~(\ref{fig.noise}). The total interferometer noise PSD is
\begin{eqnarray}
    S_{\mathrm{s}}&=& \bigg(10\,\frac{\mathrm{pm}}{\sqrt{\mathrm{Hz}}}\bigg)^2 \left[1  + \left(\frac{2\,\mathrm{mHz}}{f}\right)^{4}\right]. \label{eq.intro.2a}
\end{eqnarray}
The TM acceleration noise is
\begin{eqnarray}
    S_{\delta}&=& \bigg(\frac{3}{\omega^{2}}\,\frac{\mathrm{fm/s^{2}}}{\sqrt{\mathrm{Hz}}}\bigg)^2 \left[1  + \left(\frac{0.4\,\mathrm{mHz}}{f}\right)^{2}\right] \left[1  + \left(\frac{f}{8\,\mathrm{mHz}}\right)^{4}\right],
    \label{eq.TM_noise} 
\end{eqnarray}
and the DWS noise is modeled as
\begin{equation}\label{eq.noise.dws}
    S_{\rm n_{\phi,\eta}}=\frac{(50\,\frac{\rm nrad}{\sqrt{\rm Hz}})^{2}}{M^{2}}\left[1 + \left(\frac{2\,{\rm mHz}}{f}\right)^{4}\right]
\end{equation}
where $M$=300 is telescope and optical bench beam magnification. 

\subsection{Jitter signals \label{app.jitt.sign}}

Here we show the equations for the curves given in Fig.~(\ref{fig.jitter}) for both cases. Case A signals for SC and MOSA are
\begin{eqnarray}
S_{(\phi,\eta,\theta)_{\rm SC}}&=&(10\,{\rm nrad}/\sqrt{\rm Hz})^{2}\left[1 + \left(\frac{0.8\,{\rm mHz}}{f} \right)^4\right] \\
S_{\phi_{\rm MOSA}}&=&(10\,{\rm nrad}/\sqrt{\rm Hz})^{2}\left[1 + \left(\frac{0.8\,{\rm mHz}}{f} \right)^4\right] \\
S_{\eta_{\rm MOSA}}&=&(10\,{\rm nrad}/\sqrt{\rm Hz})^{2}\left[1 + \left(\frac{0.5\,{\rm mHz}}{f} \right)^4\right].
\end{eqnarray}
Case B signals for SC and MOSA are
\begin{eqnarray} \label{eq:LISASim_jitter1}
S_{(\phi,\eta)_{\rm SC}} &=& (40\,{\rm frad}/\sqrt{\rm Hz})^{2} \left[1 + \left(\frac{f}{f_1}\right)^2 \right]^{2}\left[1 + \frac{f(f + \alpha^2 f - 2\alpha^2 f_2)}{\alpha^2 f_2^2}\right]^{-2} \nonumber \\
&&\quad \times \left[1 + \frac{f(f + \alpha^2 f - 2\alpha^2 f_3}{\alpha^2 f_3^2}\right]^{-2} \left[1 + \left(\frac{f}{f_4}\right)^2\right]^{-2} \\
\label{eq:LISASim_jitter2} 
S_{\theta_{\rm SC}} &=& (4\,{\rm frad}/\sqrt{\rm Hz})^{2} \left[1 + \left(\frac{f}{f_1}\right)^2 \right]^{2}\left[1 + \frac{f(f + \alpha^2 f - 2\alpha^2 f_2)}{\alpha^2 f_2^2}\right]^{-2} \nonumber \\
&& \quad \times \left[1 + \frac{f(f + \alpha^2 f - 2\alpha^2 f_3}{\alpha^2 f_3^2}\right]^{-2} \left[1 + \left(\frac{f}{f_4}\right)^2\right]^{-2} \\
\label{eq:LISASim_jitter3} 
S_{\phi_{\rm MOSA}} &=& (10\,{\rm mrad}/\sqrt{\rm Hz})^{2} \left[\frac{(f^2 + f_5^2)^2}{(1 + f^2)f_5^2} \right]^2\\
\label{eq:LISASim_jitter4}
S_{\eta_{\rm MOSA}} &=& (1\,{\rm mrad}/\sqrt{\rm Hz})^{2} \left[\frac{(f^2 + f_5^2)^2}{(1 + f^2)f_5^2} \right]^2,
\end{eqnarray}
where the frequency terms are given in Table~\ref{table:LISASim}. 
\begin{table}[h!]
\begin{center}
\begin{tabular}{lccccc}
\toprule
$\alpha$ & $f_1$ & $f_2$ & $f_3$ & $f_4$ & $f_5$ \\
0.7 & \num{2e-5} & 0.01 & 0.05 & 8 & \num{1e-4} \\
\bottomrule
\end{tabular}
 \caption{Frequency values for Eqs.~(\ref{eq:LISASim_jitter1})-(\ref{eq:LISASim_jitter4}).}
 \label{table:LISASim}
\end{center}
\end{table}
\section{Time Delay Interferometry \label{app.TDI2p0}}
The main text assumes some approximations of the following equations for brevity's sake. These are the full equations for TDI 2.0, starting with the time-delay combinations of the interferometer signals, as seen in ~\cite{TDI_otto}, the $X$ combination is:
\begin{eqnarray}\label{eq.TDI2_full}
    X &=& \xi_{1'} + \xi_{3:2'}+ \xi_{1:22'}+\xi_{2':322'}+ \xi_{1:3'322'}+\xi_{2':33'322'}+\xi_{1':3'33'322'}+\xi_{3:2'3'33'322'} \nonumber \\
&&-(\xi_{1}+\xi_{2':3}+\xi_{1':3'3}+\xi_{3:2'3'3}+\xi_{1':22'3'3}+\xi_{3:2'22'3'3}+\xi_{1:22'22'3'3}+\xi_{2':322'22'3'3})
\end{eqnarray}
and assuming MOSA$_{1}$ contains the primary laser with transponder scheme, it reduces to 
\begin{eqnarray} 
X &=& \xi_{1'} + \xi_{1:22'}+ \xi_{1:3'322'}+\xi_{1':3'33'322'}
-(\xi_{1}+\xi_{1':3'3}+\xi_{1':22'3'3}+\xi_{1:22'22'3'3}).
\end{eqnarray}

Our calculations are primarily in the frequency domain, so presented here is TDI and some noise sources in the Fourier domain. 
\begin{eqnarray}
\hspace*{-0.9cm}\widetilde{X}^{\rm TTL} &=&
    +({\bf k}_{1}^{\rm RX} + {\bf k}_{1}^{\rm TX}e^{-i\omega(\tau_{3}+\tau_{3'})})\widetilde{\mathbf \Psi}_{1}\{-1+e^{-i\omega(\tau_{2}+\tau_{2'})}+e^{-i\omega(\tau_{3}+\tau_{3'}+\tau_{2}+\tau_{2'})} -
    e^{-i\omega(2(\tau_{2}+\tau_{2'})+\tau_{3}+\tau_{3'})}\}   \nonumber \\
    &&-({\bf k}_{1'}^{\rm RX} + {\bf k}_{1'}^{\rm TX}e^{-i\omega(\tau_{2}+\tau_{2'})})\widetilde{\mathbf \Psi}_{1'}\{-1 + e^{-i\omega(\tau_{3}+\tau_{3'})} +  e^{-i\omega(\tau_{3}+\tau_{3'}+\tau_{2}+\tau_{2'})} - e^{-i\omega(2(\tau_{3}+\tau_{3'})+\tau_{2}+\tau_{2'})}\}  \nonumber \\
    &&+({\bf k}_{2'}^{\rm RX}+{\bf k}_{2'}^{\rm TX})\widetilde{\mathbf \Psi}_{2'}e^{-i\omega\tau_{3}}\{-1 + 
    e^{-i\omega(\tau_{2}+\tau_{2'})} + e^{-i\omega(\tau_{3}+\tau_{3'}+\tau_{2}+\tau_{2'})} - e^{-i\omega(2(\tau_{2}+\tau_{2'})+\tau_{3}+\tau_{3'})}\}   \nonumber \\
    &&-({\bf k}_{3}^{\rm RX}+{\bf k}_{3}^{\rm TX})\widetilde{\mathbf \Psi}_{3}e^{-i\omega\tau_{2'}}\{-1 + e^{-i\omega(\tau_{3}+\tau_{3'})} -
    e^{-i\omega(\tau_{3}+\tau_{3'}+\tau_{2}+\tau_{2'})} -
    e^{-i\omega(2(\tau_{3}+\tau_{3'})+\tau_{2}+\tau_{2'})}\}
\end{eqnarray}
The $Y$ combination can be obtained from $X$ by cyclic permutation, i.e.,
\begin{eqnarray}
\label{eq.Y_2p0}
\hspace*{-0.9cm}\widetilde{Y}^{\rm TTL} &=& 
    +({\bf k}_{2}^{\rm RX}+{\bf k}_{2}^{\rm TX}e^{-i\omega(\tau_{1}+\tau_{1'})})\widetilde{\mathbf \Psi}_{2}\{
    -1 + e^{-i\omega(\tau_{3}+\tau_{3'})} + e^{-i\omega(\tau_{1}+\tau_{1'}+\tau_{3}+\tau_{3'})} - 
    e^{-i\omega(2(\tau_{3}+\tau_{3'})+\tau_{1}+\tau_{1'})} 
    \}  \nonumber \\
    &&-({\bf k}_{2'}^{\rm RX}+{\bf k}_{2'}^{\rm TX}e^{-i\omega(\tau_{3}+\tau_{3'})})\widetilde{\mathbf \Psi}_{2'}\{
    -1 + e^{-i\omega(\tau_{1}+\tau_{1'})} + e^{-i\omega(\tau_{1}+\tau_{1'}+\tau_{3}+\tau_{3'})} -
    e^{-i\omega(2(\tau_{1}+\tau_{1'})+\tau_{3}+\tau_{3'})} 
    \} \nonumber \\
    &&+({\bf k}_{3'}^{\rm RX}+{\bf k}_{3'}^{\rm TX})\widetilde{\mathbf \Psi}_{3'}e^{-i\omega\tau_{1}}\{
    -1 +  e^{-i\omega(\tau_{3}+\tau_{3'})} + e^{-i\omega(\tau_{1}+\tau_{1'}+\tau_{3}+\tau_{3'})} -
    e^{-i\omega(2(\tau_{3}+\tau_{3'})+\tau_{1}+\tau_{1'})} 
    \} \nonumber \\
    &&-({\bf k}_{1}^{\rm RX}+{\bf k}_{1}^{\rm TX})\widetilde{\mathbf \Psi}_{1}e^{-i\omega\tau_{3'}}\{
    -1 + e^{-i\omega(\tau_{1}+\tau_{1'})} + e^{-i\omega(\tau_{1}+\tau_{1'}+\tau_{3}+\tau_{3'})} -
    e^{-i\omega(2(\tau_{1}+\tau_{1'})+\tau_{3}+\tau_{3'})} 
    \} 
    \label{eq.Y_TTL_2p0}
\end{eqnarray}
and similarly to obtain $Z$ from $Y$.
Next we derive the interferometer noise in TDI X 2.0
\begin{eqnarray}
\hspace*{-0.5cm}S_{\rm IFO}^{X} &=&
(S_{\rm s_{1}}+S_{\rm s_{2'}})|-1 + e^{-i\omega(\tau_{2}+\tau_{2'})}+e^{-i\omega(\tau_{3}+\tau_{3'}+\tau_{2}+\tau_{2'})}-e^{-i\omega(2(\tau_{2}+\tau_{2'})+ \tau_{3}+\tau_{3'})}|^{2} 
\nonumber \\
&&+ (S_{\rm s_{1'}}+S_{\rm s_{3}})|+1 - e^{-i\omega(\tau_{3}+\tau_{3'})}-e^{-i\omega(\tau_{3}+\tau_{3'}+\tau_{2}+\tau_{2'})}+e^{-i\omega(2(\tau_{3}+\tau_{3'})+ \tau_{2}+\tau_{2'})}|^{2}  
\end{eqnarray}
where $S_{{\rm s}_{i}}$ is the one-link interferometer noise. 
$S_{\rm IFO}^{Y}$ and $S_{\rm IFO}^{Z}$ can be derived by cyclic permutation from $X$ and $Y$, respectively. The TM acceleration noise after TDI is
\begin{eqnarray}
&&\hspace*{-0.5cm}S_{\rm TM}^{X} = \nonumber \\
&&S_{\delta_{1}}|-1+e^{-i\omega(\tau_{2}+\tau_{2'})}-e^{-i\omega(\tau_{3'}+\tau_{3})}+2e^{-i\omega(\tau_{2}+\tau_{2'}+\tau_{3}+\tau_{3'})}+e^{-i\omega(2(\tau_{3'}+\tau_{3})+\tau_{2}+\tau_{2'})}-e^{-i\omega(2(\tau_{2}+\tau_{2'})+\tau_{3'}+\tau_{3})} \nonumber \\
&&\quad - e^{-i\omega(2(\tau_{2}+\tau_{2'}+\tau_{3}+\tau_{3'}))}|^{2}+\nonumber \\
&&S_{\delta_{1'}}|1+e^{-i\omega(\tau_{2}+\tau_{2'})}-e^{-i\omega(\tau_{3'}+\tau_{3})}-2e^{-i\omega(\tau_{2}+\tau_{2'}+\tau_{3}+\tau_{3'})}+e^{-i\omega(2(\tau_{3'}+\tau_{3})+\tau_{2}+\tau_{2'})}-e^{-i\omega(2(\tau_{2}+\tau_{2'})+\tau_{3'}+\tau_{3})} \nonumber \\
&&\quad + e^{-i\omega(2(\tau_{2}+\tau_{2'}+\tau_{3}+\tau_{3'}))}|^{2}+\nonumber \\
&&4S_{\delta_{3}}|+1 - e^{-i\omega(\tau_{3'}+\tau_{3})}-e^{-i\omega(\tau_{2'}+\tau_{2}+\tau_{3'}+\tau_{3})}+
e^{-i\omega(2(\tau_{3}+\tau_{3'})+\tau_{2}+\tau_{2'})}|^{2} +\nonumber \\
&&4S_{\delta_{2'}}|-1 + e^{-i\omega(\tau_{2}+\tau_{2'})}+e^{-i\omega(\tau_{3}+\tau_{3'}+\tau_{2}+\tau_{2'})}-
e^{-i\omega(2(\tau_{2}+\tau_{2'}) +\tau_{3}+\tau_{3'})}|^{2} 
\end{eqnarray}
where $S_{\delta_{i}}$ is the acceleration noise of the a single TM.
$S_{\rm n_{TM\,all}}^{Y}$ and $S_{\rm n_{TM\,all}}^{Z}$ can be derived by cycling permutation. The DWS noise in $\phi$ after propagation through TDI is
\begin{eqnarray}
  S_{{\rm n}_{\phi}}^{X} &=&[(c_{1}^{\rm RX})^{2}+(c_{1}^{\rm TX})^{2}+2c_{1}^{\rm RX}c_{1}^{\rm TX}\cos{\omega(\tau_{3}+\tau_{3'})}]S_{\rm n_{\phi_{1}}} \nonumber \\
  &&\quad \times |-1+e^{-i\omega(\tau_{2}+\tau_{2'})}+e^{-i\omega(\tau_{3'}+\tau_{3}+\tau_{2}+\tau_{2'})} -
    e^{-i\omega(2(\tau_{2}+\tau_{2'})+\tau_{3'}+\tau_{3})}|^{2}  \nonumber \\
    &&+[(c_{1'}^{\rm RX})^{2}+(c_{1'}^{\rm TX})^{2}+2c_{1'}^{\rm RX}c_{1'}^{\rm TX}\cos{\omega(\tau_{2}+\tau_{2'})}]S_{\rm n_{\phi_{1'}}} \nonumber \\
    &&\quad \times
    |+1 - e^{-i\omega(\tau_{3'}+\tau_{3})}-e^{-i\omega(\tau_{2}+\tau_{2'}+\tau_{3'}+\tau_{3})}+e^{-i\omega(2(\tau_{3'}+\tau_{3})+\tau_{2}+\tau_{2'})}|^{2} \nonumber \\
    &&+(c_{2'}^{\rm RX}+c_{2'}^{\rm TX})^{2}S_{\rm n_{\phi_{2'}}}|-1+e^{-i\omega(\tau_{2}+\tau_{2'})}+e^{-i\omega(\tau_{3}+\tau_{3'}+\tau_{2}+\tau_{2'})}-e^{-i\omega(2(\tau_{2}+\tau_{2'})+\tau_{3}+\tau_{3'})}|^{2} \nonumber \\
    &&+ (c_{3}^{\rm RX}+c_{3}^{\rm TX})^{2}S_{\rm n_{\phi_{3}}}|+1-e^{-i\omega(\tau_{3}+\tau_{3'})}-e^{-i\omega(\tau_{2'}+\tau_{2}+\tau_{3'}+\tau_{3})}+e^{-i\omega(2(\tau_{3'}+\tau_{3})+\tau_{2}+\tau_{2'})}|^{2}
\end{eqnarray}
and for $\eta$ one replaces $c$ with $d$ and $S_{\rm n_{\phi}}^{X}$ for $S_{\rm n_{\eta}}^{X}$. The total DWS noise is
\begin{equation}\label{eq.GW.app}
    S_{\rm DWS}^{X}=S_{\rm n_{\phi}}^{X} + S_{\rm n_{\eta}}^{X}
\end{equation}
and similarly for $Y$ and $Z$ after cyclic permutation of the indices. Finally, for the gravitational-wave background, we model it roughly based on the Mock LISA Data Challenges (MLDC) data set~\cite{Babak_2008}:                                         
\begin{equation}
    S_{\rm h}=4\pi^{2}10^{-44}/\omega^{2}
\end{equation}

\section{Residuals after TTL subtraction}\label{app.residual}
The TTL noise residual making no assumptions on the time-delays is (for $\phi$)
%
%
\begin{eqnarray}
S_{\rm res, \phi}^{X} &=& (S_{\phi_{{\rm SC}_{1}}} + S_{\phi_{{\rm M}_{1}}})\big[ \big( \sigma_{11}^2 + \sigma_{22}^2 + 2\sigma_{12}\cos{\omega(\tau_3 + \tau_{3'}})\big] |H_1|^2 \nonumber \\
&& + (S_{\phi_{{\rm SC}_{1}}} + S_{\phi_{{\rm M}_{1'}}}) \big[ \sigma_{33}^2 + \sigma_{44}^2 + 2\sigma_{34}\cos{\omega(\tau_2 + \tau_{2'}})\big] |H_2|^2  \nonumber \\
&& - 2 S_{\phi_{{\rm SC}_{1}}} \big[\sigma_{13} \Re{\{H_1 H_2^*\}} + \sigma_{14} \Re{\{H_1 (H_2 e^{-i\omega (\tau_2 + \tau_{2'})})^*\}} \nonumber \\
&& \quad + \sigma_{23} \Re{\{H_1 (H_2 e^{i\omega (\tau_3 + \tau_{3'})})^*\}}  + \sigma_{24} \Re{\{H_1 (H_2 e^{i\omega (\tau_3 + \tau_{3'} -\tau_2 - \tau_{2'})})^*\}}\big] \nonumber \\
&& + (S_{\phi_{{\rm SC}_{2}}}+S_{\phi_{{\rm M}_{2'}}}) \sigma_{55}^2 |H_1|^2 + (S_{\phi_{{\rm SC}_{3}}}+S_{\phi_{{\rm M}_{3}}}) \sigma_{66}^2 |H_2|^2 \nonumber \\
&& + S_{{\rm n}_{\phi}}^{X}
\end{eqnarray}
%
%
where
\begin{eqnarray}
    H_1 &=& -1 + e^{-i\omega(\tau_{2}+\tau_{2'})} + e^{-i\omega(\tau_{3}+\tau_{3'}+\tau_{2}+\tau_{2'})} -
    e^{-i\omega(2(\tau_{2}+\tau_{2'})+\tau_{3}+\tau_{3'})} \\
    H_2 &=& -1 + e^{-i\omega(\tau_{3}+\tau_{3'})} + e^{-i\omega(\tau_{3}+\tau_{3'}+\tau_{2}+\tau_{2'})} -
    e^{-i\omega(2(\tau_{3}+\tau_{3'})+\tau_{2}+\tau_{2'})}.
\end{eqnarray}
The residuals for $\eta$ are calculated by replacing $\phi$ by $\eta$ and 
by changing the covariance elements to $\sigma_{i+6,j+6}$ with $i,j=1,2,..6$. 
As previously, the residuals for $Y$ and $Z$ can be obtained from cyclic permutation of the indices. 
\bibliographystyle{unsrt}

\bibliography{references.bib}

\end{document}